\documentclass
[twocolumn,aps,prc,amsmath,showpacs,floatfix,nofootinbib
,superscriptaddress
]{revtex4-1}

\usepackage{CJK}                     
\usepackage[dvips]{graphicx}
\usepackage{bm}                      
\usepackage{xcolor,soul}\sethlcolor{green}
\usepackage{mathptmx}                
\usepackage{dcolumn}                 

\def\be{\begin{equation}}  \def\ee{\end{equation}}
\def\bea{\begin{eqnarray}} \def\eea{\end{eqnarray}}
\def\bal#1\eal{\begin{align}#1\end{align}}
\def\bse#1\ese{\begin{subequations}#1\end{subequations}}
\def\non{\nonumber}
\def\ra{\rightarrow}
\renewcommand\Re{\operatorname{Re}}
\renewcommand\Im{\operatorname{Im}}

\def\eps{\varepsilon}
\def\rv{\bm{r}} \def\Jv{\bm{J}} \def\Wv{\bm{W}}
\def\la{\Lambda} \def\la{Y} \def\lala{\Lambda\Lambda}
\def\xim{$\Xi^-$}
\def\ron{\rho_N} \def\tn{\tau_N}
\def\rol{\rho_\la} \def\tl{\tau_\la} \def\ml{m_\la}
\def\xbe{$^{12}_{\hskip0.27em\Xi}$Be} \def\xbem{\text{\xbe}}
\def\xic{$^{15}_{\hskip0.27em\Xi}$C}  
\def\xib{$^{13}_{\hskip0.27em\Xi}$B}  
\def\xip{$^{14}$N~(g.s.)$+\Xi^-(1p)$}
\def\xis{$^{14}$N~(g.s.)$+\Xi^-(1s)$}
\def\xicp{$^{15}_{\hskip0.05em\Xi p}$C}
\def\xbes{$^{12}_{\hskip0.05em\Xi s}$Be}
\def\lbe{$^{10}_{\hskip0.29em\Lambda}$Be}
\def\bxi{B_{\Xi^-}}
\def\xin{$\Xi N$}

\begin{document}

\title{Mean field approaches for $\bm\Xi^-$ hypernuclei
and current experimental data}

\begin{CJK*}{GB}{gbsn} 

\author{T. T. Sun}
\affiliation{
RIKEN Nishina Center, Wako 351-0198, Japan}
\affiliation{
School of Physics and Engineering, Zhengzhou University,
Zhengzhou 450001, China}

\author{E. Hiyama}
\email{hiyama@riken.jp}
\affiliation{
RIKEN Nishina Center, Wako 351-0198, Japan}

\author{H. Sagawa}
\affiliation{
RIKEN Nishina Center, Wako 351-0198, Japan}
\affiliation{
Center for Mathematics and Physics, University of Aizu,
Aizu-Wakamatsu, Fukushima 965-8560, Japan}

\author{H.-J. Schulze}
\affiliation{
INFN Sezione di Catania, Dipartimento di Fisica, Universit\'a di Catania,
Via Santa Sofia 64, I-95123 Catania, Italy}

\author{J. Meng}
\affiliation{
School of Physics and State Key Laboratory of Nuclear Physics and Technology,
Peking University, Beijing 100871, China}
\affiliation{
School of Physics and Nuclear Energy Engineering, Beihang University,
Beijing 100191, China}


\begin{abstract}
Motivated by the recently observed hypernucleus (Kiso event)
\xic\ ($^{14}$N$+\Xi^-$),
we identify the state of this system theoretically
within the framework of the relativistic-mean-field
and Skyrme-Hartree-Fock models.
The \xin\ interactions are constructed to reproduce
the two possibly observed $\Xi^-$ removal energies,
$4.38\pm 0.25$~MeV or $1.11\pm 0.25$~MeV.
The present result is preferable to be \xip,
corresponding to the latter value.
\end{abstract}

\pacs{
 21.80.+a, 
 13.75.Ev, 
 21.60.Jz, 
 21.10.Dr 
}

\maketitle

\end{CJK*}

\section{Introduction}

One of the goals of hypernuclear physics is to obtain useful information on
the baryon-baryon interactions in a unified way,
which is important in particular for astrophysical applications \cite{astro}.
However, hyperon-nucleon ($Y\!N$) scattering data are very limited due to the
difficulty of $Y\!N$ scattering experiments,
and there are no $YY$ scattering data at all.
Thus, the existing $Y\!N$ and $YY$ potential models have a lot of ambiguity,
and in order to constrain them better
it is important to study the structure of hypernuclei,
such as single-, double-$\Lambda$ hypernuclei,
and $\Xi$ hypernuclei.

For the $\Lambda N$ interaction,
rich experimental information on $\Lambda$ hypernuclei
is available \cite{hashimoto},
in particular accurate measurements of $\gamma$-ray spectra have been performed
systematically \cite{gamma},
and used to extract information on the spin-dependent components
of the $\Lambda N$ interactions through detailed analyses of
hypernuclear structure,
using the shell model \cite{Millener} or cluster models
with the Gaussian Expansion Method \cite{Hiyama2009},
for example.

For the study of interactions within the strangeness  $S=-2$ sector,
the observed $\lala$ bond energies of double-$\Lambda$ hypernuclei
are currently the only reliable source of information on
the $\lala$ interaction.
In this regard, we stress the importance of
the observation of the double-$\Lambda$ hypernucleus
$^{\ \ 6}_{\lala}$He
(NAGARA event) in the KEK-E373 experiment \cite{NAGARA}.
Further analysis of this experiment is still in progress.

Regarding the \xin\ interaction,
the few current experimental data indicate that
the $\Xi$-nucleus interactions are attractive.
One example is the observed spectrum
of the $(K^-,K^+)$ reaction on a $^{12}$C target to produce \xbe,
where the cross section for $\Xi^-$ production in the threshold region
was interpreted by assuming a $\Xi^-$-nucleus Woods-Saxon (WS) potential with
a depth of about 14~MeV \cite{khaustov}.
Using this assumption,
a cluster model calculation
predicted values of about 5~MeV
for the ground-state \xim\ binding (removal) energy
$\bxi \equiv E(\xbem)-E(^{11}\text{B})$ with Coulomb interaction
for \xbe\ and 2.2~MeV without \cite{hiyama},
while an AMD approach \cite{amd}
yielded slightly lower values of about 3--5.5~MeV,
using the same \xin\ interactions.

Other observed data on the $\Xi$ hypernucleus
\xib\ ($^{12}{\rm C}+\Xi^-$)
were obtained by emulsion data \cite{Twin,yama}.
The reported $\Xi^-$ binding energies are
$3.70^{+0.18}_{-0.19}$~MeV,
$0.62^{+0.18}_{-0.19}$~MeV, and
$2.66 ^{+0.18}_{-0.19}$~MeV,
where the second value was expected to be consistent with
a decay from this system in the $2P$ state.
However, there is also the possibility that the observed event
was a decay from an atomic $3D$ state.
Therefore, it is hard to confirm that this event was an observation
of a strongly bound $\Xi$ hypernucleus.

In 2015, analysis of the KEK-E373 experiment provided the first clear
evidence of the bound $\Xi^-$ hypernucleus \xic\ \cite{nakazawa},
produced in the reaction
$\Xi^- + \text{$^{14}$N} \ra$ \xic\ $\ra$ \lbe\ $+\;\text{$^5_\Lambda$He}$.
This data is called ``Kiso'' event.
Two possible \xim\ binding energies are interpreted experimentally:
(1)
One is
$\bxi \equiv E$(\xic)$-E(^{14}\text{N})
= 4.38\pm0.25\,\text{MeV}$,
which was deduced under the assumption that both hypernuclei \xic\ and \lbe\
involved in the reaction were produced in their ground states.
In this case, the hypernucleus \xic\ is considered to be in the state \xis.
(2)
Another possible binding energy is
$\bxi = 1.11\pm0.25\,\text{MeV}$,
if \lbe\ was left in an excited  state.
Recently JLab reported the energy spectra of \lbe\ using the
$(e,e'K^{+})$ reaction~\cite{10LBe}.
When the energy of the excited state in \lbe\ is taken into account,
the $\Xi^-$ binding energy is interpreted around 1.11~MeV,
and the observed \xic\ in the Kiso event is in the state
\xip\ \cite{private}.

The Kiso event is important in the sense that
it confirms that the \xin\ interaction is attractive.
Now we have the following questions:
(i) Can we identify the state of the Kiso event theoretically, that is,
the event is \xis\ or \xip?
(ii) How much attraction do we need to
reproduce the observed $\Xi^-$ binding energies of the event?
(iii) Can we reproduce the old \xbe\ data in Ref.~\cite{khaustov,hiyama}
using the same \xin\ interaction to reproduce the Kiso event?

To answer these questions,
we adopt the relativistic mean field (RMF) and Skyrme Hartree-Fock (SHF) models
to study the Kiso event by employing effective interactions,
which will be fitted to reproduce the experimental \xim\ binding energy of \xic,
and with those interactions we investigate
whether a consistent theoretical description for
\xic\ and \xbe\ is possible or not.

The mean-field theory is a powerful theoretical approach,
which can be globally applied from light to heavy (hyper)nuclei \cite{RMFlambda}.
It should be stressed that this approach has also been employed successfully
for $\Lambda$ hypernuclei with $A\sim10$,
which are relatively light systems
\cite{ray,shflight,hypsky,rmflight,RMFlambda}.
In the present work,
we study $\Xi$ hypernuclei with $A=12$ and 15 and focus on the states
$^{11}$B~(g.s.)$+\Xi(1s)$ for \xbe\ and
$^{14}$N~(g.s.)$+\Xi^-(1s,1p)$ for \xic,
and their hyperon separation energy,
\be
 B_Y = E([n,p,Y]) - E([n,p,-]) \:.
\ee
It should be noted that the core nuclei of $^{11}$B and $^{14}$N
are compact shell structures and then it is not expect to
have any dynamical contraction of the core by addition of a hyperon.
This phenomena was already pointed out in Ref. \cite{Hiyama97}.
Thus, for this observable one can expect that a major part of an inaccurate description
of the common nuclear core $[n,p]$ cancels out,
as well as that other uncertainties such as
center-of-mass, pairing, deformation corrections etc.,
become much less relevant.
The removal energy then depends predominantly on the phenomenological
$Y\!N$ interaction parameters that we adjust to the data,
i.e., the hyperon-nucleus mean field.
Therefore it is expected that we can safely interpret
the state of the Kiso event theoretically.

It should also be noted that the $\Xi^-$ hypernuclei decay into
double-$\Lambda$ hypernuclei by the \xin--$\lala$ coupling.
Therefore, the \xin\ interaction should have an imaginary part
to represent the decay width.
However, since we have no experimental information on this coupling
by Refs.~\cite{khaustov,nakazawa},
here the imaginary part is omitted.

This article is organized as follows:
In Sec.~\ref{s:theo},
the theoretical methods and interactions are briefly described.
The numerical results and corresponding discussions for \xic\ and \xbe\
are presented in Sec.~\ref{s:res}.
A summary is given in Sec.~\ref{s:end}.

\section{Models and interactions}
\label{s:theo}
\subsection{Relativistic Mean Field Model}
\label{Subsec:RMF}

The starting point of the meson-exchange RMF
model for hypernuclei is the covariant Lagrangian density
\be
 \mathcal{L} = \mathcal{L}_N + \mathcal{L}_Y \:,
\ee
where $\mathcal{L}_N$ is the standard RMF Lagrangian density for the
nucleons \cite{rmf,PPNP2006MengJ},
and $\mathcal{L}_{Y}$ is the Lagrangian density for the hyperons
\cite{PRC1994Mares.49.2472},
in which the couplings with the scalar $\sigma$, vector $\omega_\mu$,
vector-isovector ${\bm\rho}_\mu$ mesons, and the photon $A_\mu$
are included.
For the charged $Y=\Xi$ hyperon with isospin 1/2,
the Lagrangian density $\mathcal{L}_Y$ reads
\bea
 && \mathcal{L}_Y = \overline{\psi}_\Xi \Big[
 i\gamma^\mu\partial_\mu - m_\Xi -
 g_{\sigma\Xi}\sigma - g_{\omega\Xi}\gamma^\mu\omega_\mu
\\&&\hskip5mm
 - g_{\rho\Xi}\gamma^\mu{\bm\tau}_\Xi\cdot{\bm\rho}_\mu
 - e\gamma^\mu\frac{\tau_{\Xi,3}-1}{2}A_\mu
 - \frac{f_{\omega\Xi}}{2m_\Xi}
 \sigma^{\mu\nu}\partial_\nu\omega_\mu
 \Big]\psi_\Xi \:,
\nonumber
\label{EQ:Lagrangian}
\eea
where $m_\Xi$
is the mass of the $\Xi$ hyperon,
$g_{\sigma\Xi}$, $g_{\omega\Xi}$, and $g_{\rho\Xi}$
are the coupling constants of the $\Xi$ hyperon with the
$\sigma$, $\omega$, and $\rho$ mesons, respectively,
and $\tau_{\Xi,3}$ is the third component of the isospin vector
${\bm\tau}_\Xi$
($+1$ for the neutral $\Xi^0$ and $-1$ for the negatively charged $\Xi^-$).
The last term in $\mathcal{L}_Y$ is the tensor coupling
with the $\omega$ field.
For the studies of $\Lambda$ hypernuclei with the RMF model,
see \cite{RMFlambda} and references therein.

For a system with time-reversal symmetry,
the space-like components of the vector fields vanish,
only leaving the time components $\omega_0$, ${\bm\rho}_0$, $A_0$.
Furthermore, one can assume that in all nuclear applications,
the hyperon single-particle (s.p.) states do not mix isospin, i.e.,
the s.p.~states are the eigenstates of $\tau_{\Xi,3}$,
and therefore only the third component of the $\rho_0$ meson field,
$\rho_{0,3}$, survives.

With the mean-field and no-sea approximations,
the s.p.~Dirac equations for baryons
and the Klein-Gordon equations for mesons and photon
can be obtained by the variational procedure.
In the spherical case, the Dirac spinor can be expanded as
\be
 \psi_{n\kappa m}({\bm r}) =
 \left(\begin{array}{c}
   iG_{n\kappa}(r) \\
    F_{n\kappa}(r) {\bm\sigma}\cdot{\hat{\bm r}} \\
 \end{array}\right) \frac{Y_{jm}^l(\theta,\phi)}{r} \:,
\label{EQ:RWF}
\ee
where $G_{n\kappa}(r)/r$ and $F_{n\kappa}(r)/r$
are the radial wave functions for the upper and lower components,
$Y_{jm}^l(\theta,\phi)$ are the spinor spherical harmonics,
and the quantum number $\kappa$ is defined by the angular momenta $(l,j)$
as $\kappa=(-1)^{j+l+1/2}(j+1/2)$.

The Dirac equation for the radial wave functions of the $\Xi$ hyperon is
\be
 \left(\begin{array}{cc}
  V+S                             & -\frac{d}{dr}+\frac{\kappa}{r}+T \\
  \frac{d}{dr}+\frac{\kappa}{r}+T & V-S-2m_\Xi                       \\
 \end{array}\right)
 \left(\begin{array}{c}
  G_{n\kappa}^\Xi \\
  F_{n\kappa}^\Xi \\
 \end{array}\right)
 = e_{n\kappa}^\Xi
 \left(\begin{array}{c}
  G_{n\kappa}^\Xi \\
  F_{n\kappa}^\Xi \\
 \end{array}\right) \:,
\label{EQ:RDirac}
\ee
where $e_{n\kappa}^\Xi$ is the s.p.~energy,
and
\bse
\bea
 S &=& g_{\sigma\Xi}\sigma \:,
\\\label{e:V2}
 V &=& g_{\omega\Xi}\omega_0 + g_{\rho\Xi}\tau_{\Xi,3}\rho_{0,3}
 + e\frac{\tau_{\Xi,3}-1}{2}A_0 \:,
\\
 T &=& -\frac{f_{\omega\Xi}}{2m_\Xi}\partial_r\omega_0 \:
\eea
\label{e:svt}%
\ese%
are the scalar, vector, and tensor potentials, respectively.

The meson and photon fields satisfy the radial Laplace equations
\be
 \left( -\frac{d^2}{dr^2} - \frac{2}{r}\frac{d}{dr} + m_\phi^2 \right)\phi
 = S_\phi
\ee
with the source terms
\be
 S_\phi =
 \left\{\begin{array}{ll}{\displaystyle
 -g_\sigma\rho_s - g_{\sigma\Xi}\rho_{s\Xi} - g_2\sigma^2 - g_3\sigma^3} \:,
\phantom{\Big|}& \hbox{for~$\sigma$} \:,
\\{\displaystyle
 g_\omega\rho_v + g_{\omega\Xi}\rho_{v\Xi} +
 \frac{f_{\omega\Xi}}{2m_\Xi} \partial_ij_{T\Xi}^{0i} - c_3\omega_0^3} \:,
\phantom{\Big|}& \hbox{for~$\omega_0$} \:,
\\{\displaystyle
 g_\rho\rho_3 + g_{\rho\Xi}\rho_{3\Xi} - d_3\rho_{0,3}^3} \:,
\phantom{\Big|}& \hbox{for~$\rho_{0,3}$} \:,
\\{\displaystyle
 e\rho_c + e\rho_{c\Xi}} \:,
\phantom{\Big|}& \hbox{for~$A_0$} \:,
\end{array}\right.
\label{EQ:LaplaceEq}
\ee
where $m_\phi$ are the meson masses for
$\phi=\sigma, \omega_0, \rho_{0,3}$
and zero for the photon,
$g_\sigma$, $g_\omega$, $g_\rho$, $g_2$, $g_3$, $c_3$, and $d_3$
are the parameters for the nucleon-nucleon ($NN$) interaction
in the Lagrangian density
$\mathcal{L}_N$ \cite{PPNP2006MengJ},
$\rho_s(\rho_{s\Xi})$, $\rho_v(\rho_{v\Xi})$, $\rho_3(\rho_{3\Xi})$,
and $\rho_c(\rho_{c\Xi})$ are the radial scalar, baryon, isovector,
and charge densities for the nucleons (hyperons), respectively, and
$j_{T\Xi}^{0i}$ is the tensor density for the $\Xi$ hyperons.

With the radial wave functions,
these densities for the $\Xi$ hyperons can be expressed as
\bse
\bea
 \rho_{s\Xi}(r) &=& \frac{1}{4\pi r^2}\sum_{k=1}^{A_\Xi}
 \left[|G_k^\Xi(r)|^2-|F_k^\Xi(r)|^2\right] \:,
\\
 \rho_{v\Xi}(r) &=& \frac{1}{4\pi r^2}\sum_{k=1}^{A_\Xi}
 \left[|G_k^\Xi(r)|^2+|F_k^\Xi(r)|^2\right] \:,
\\
 \rho_{3\Xi}(r) &=& \frac{1}{4\pi r^2}\sum_{k=1}^{A_\Xi}
 \left[|G_k^\Xi(r)|^2+|F_k^\Xi(r)|^2\right]\tau_{\Xi,3} \:,
\\
\rho_{c\Xi}(r) &=& \frac{1}{4\pi r^2}\sum_{k=1}^{A_\Xi}
 \left[|G_k^\Xi(r)|^2+|F_k^\Xi(r)|^2\right]
 \frac{\tau_{\Xi,3}-1}{2} \:,\qquad
\\
 {\bm j}_{T\Xi}^{0} &=& \frac{1}{4\pi r^2}\sum_{k=1}^{A_\Xi}
 \left[2G_k^\Xi(r)F_k^\Xi(r)\right]{\bm n} \:,
\eea%
\label{EQ:Density}%
\ese%
where $\bm n$ is the angular unit vector.
The hyperon number $A_\Xi$ can be calculated by the integral
of the baryon density $\rho_{v\Xi}(r)$ in coordinate space as
\be
 A_\Xi = \int 4\pi r^2dr\; \rho_{v\Xi}(r) \:.
\label{e:axi}
\ee
The coupled equations (\ref{EQ:RDirac})-(\ref{e:axi})
in the RMF model are solved by iteration in coordinate space.

As the translational symmetry is broken in the mean-field approximation,
a proper treatment of the center-of-mass (c.m.) motion is very important,
especially for light nuclei.
In the present calculation,
we employ the microscopic c.m.~correction as in \cite{EPJA2000Bender},
\be
 E_{\rm c.m.} = -\frac{1}{2M} \langle \widehat{\bm P}^2 \rangle \:,
\label{e:cms}
\ee
where
$M = \sum_B M_B = A M_N + A_\Xi m_\Xi$
is the total mass of the (hyper)nucleus and
$\widehat{\bm P} = \sum_B \widehat{\bm P}_B$
is the total momentum operator.
With the c.m.~correction,
the total energy for the hypernucleus in RMF is finally given as
\bea
  E_\text{tot} =&& \sum_{k=1}^A e_k + \sum_{k=1}^{A_\Xi} e_k^\Xi
 - 2\pi\int r^2dr \times
\nonumber\\
 \Bigg[
 && \phantom{+} g_\sigma\rho_s\sigma + g_{\sigma\Xi}\rho_{s\Xi}\sigma
 + \frac{1}{3}g_2\sigma^3 + \frac{1}{2}g_3\sigma^4
\nonumber\\
 && + g_\omega\rho_v\omega_0 + g_{\omega\Xi}\rho_{v\Xi}\omega_0 +
 \frac{f_{\omega\Xi}}{2m_\Xi}\partial_ij_{T\Xi}^{0i}\omega_0
 -\frac{1}{2}c_3\omega_0^4
\nonumber\\
 && + g_\rho\rho_3\rho_{0,3} + g_{\rho\Xi}\rho_{3\Xi}\rho_{0,3}
 -\frac{1}{2}d_3\rho_{0,3}^4
\nonumber\\
 && + e \rho_c A_0 + e \rho_{c\Xi}A_0 \Bigg] \: + E_{\rm c.m.} \:.
\eea

In this work, the RMF Dirac equation is solved in a box of size
$R=20~{\rm fm}$ and a step size of $0.05~{\rm fm}$.
For the $NN$ interaction,
the PK1~\cite{PRC2004Long} parameter set is used.
For the \xin\ interaction,
the scalar coupling constant $g_{\sigma\Xi}$ is adjusted
in order to reproduce the \xim\ binding energy of \xic\ in either the
ground ($s$) state $\bxi \approx 4.4\;\text{MeV}$
or the excited ($p$) state
$\bxi \approx 1.1\;\text{MeV}$ \cite{nakazawa,private}.
The vector coupling constant $g_{\omega\Xi}=g_\omega/3$
is determined from the naive quark model \cite{PPNP1984Dover.12.171},
and the tensor coupling constant $f_{\omega\Xi}=-0.4g_{\omega\Xi}$
is taken as in Refs.~\cite{PRC1975Moldauer.12.744,PRC1994Mares.49.2472}.
The vector-isovector coupling constant $g_{\rho\Xi}=g_\rho$
\cite{PRC1994Mares.49.2472}
is determined by the SU(3) Clebsch-Gordan coefficients.

\subsection{Skyrme-Hartree-Fock Model}
\label{s:shf}

We employ a model based on the one-dimensional (spherical)
self-consistent SHF method \cite{vaut,shf},
first extended to the theoretical description of $\Lambda$ hypernuclei
in Ref.~\cite{ray},
and now used for hyperons $Y=\Xi^-$ here.
The fundamental SHF local energy density functional of hypernuclear matter
is written as
\be
 \eps_\text{SHF} = \eps_N + \eps_\la \:,
\ee
and depends on the one-body densities $\rho_q$,
kinetic densities $\tau_q$,
and spin-orbit currents $\Jv_q$,
\be
  \Big[ \rho_q,\; \tau_q,\; \Jv_q \Big] =
  \sum_{i=1}^{N_q} {n_q^i} \Big[
  |\phi_q^i|^2 ,\;
  |\nabla\phi_q^i|^2 ,\;
  {\phi_q^i}^* (\nabla \phi_q^i \times \bm{\sigma})/i
 \Big] \:,
\ee
where $\phi_q^i$ ($i=1,N_q$) are the
self-consistently calculated s.p.~wave functions
of the $N_q$ occupied states for the species $q=n,p,\la$ in a hypernucleus.

The functional $\eps_N$ is the usual nucleonic part \cite{vaut,shf}
and a possible standard parametrization for the hyperonic part is
\cite{ray,hypsky}
\bal
 \eps_\la =&
{\tl\over 2\ml}
 + a_0 \rol\ron
 + a_3 \rol\ron^2 
 + a_1 \left( \rol\tn + \ron\tl \right)
\label{e:epsl}
\\&
 - a_2 \left( \rol\Delta\ron+\ron\Delta\rol \right)\!/2
\non
 - a_4 \left( \rol\nabla\cdot\Jv_N + \ron\nabla\cdot\Jv_\la \right) \:,
\eal
from which one obtains the corresponding hyperonic SHF mean fields
\bal
 V_\la &=
  a_0 \ron + a_1 \tn - a_2 \Delta\ron - a_4 \nabla\cdot\Jv_N
  + a_3 \ron^2 
\:,
\label{e:vl}
\\
 V_N^{(\la)} &=
  a_0 \rol + a_1 \tl - a_2 \Delta\rol - a_4 \nabla\cdot\Jv_\la
  + 2a_3\rol\ron \:,
\eal
and a $\la$ effective mass
\bea
  {1\over 2\ml^*} &=& {1\over 2\ml} + a_1 \ron  \:.
\label{e:mfit}
\eea
The relation to the standard $\la\!N$ Skyrme parameters $t_{0,1,2,3}^{Y\!N}$ is
\bal
 a_0 = t_0 \:,\ \
 a_1 = \frac{t_1+t_2}{4}  \:,\ \
 a_2 = \frac{3t_1-t_2}{8} \:,\ \
 a_3 = \frac{3t_3}{8}      \:.
\eal

Minimizing the total energy of the hypernucleus,
$E=\int\!d^3\rv\:\eps_\text{SHF}(\rv)$,
one arrives at the SHF Schr\"odinger equation
\bal
 & \Bigg[ \nabla \cdot {1\over2m_q^*(\rv)}\nabla - V_q(\rv) - e_qV_C(\rv)
 + i \Wv_q(\rv) \cdot \left(\nabla\times\bm{\sigma}\right)
 \Bigg] \phi_q^i(\bm{r})
\non\\
 & \quad = e_q^i \,\phi_q^i(\bm{r}) \:,
\eal
where
$V_C$ is the Coulomb field
and $\Wv_N$ the nucleonic spin-orbit mean-field \cite{shf}.
In contrast to $\Lambda$ hypernuclei,
the Coulomb interaction is very important for the light \xim\ hypernuclei
discussed here.
An approximate c.m.~correction is applied as usual \cite{shf}
by replacing the bare masses:
\be
 {1\over m_q} \rightarrow {1\over m_q} - {1\over M} \:,
\label{e:cm}
\ee
where $M=(N_n+N_p)m_N + N_\la\ml$ is the total mass of the (hyper)nucleus.
This correction is the one to be used with most
nucleonic Skyrme forces \cite{vaut,shf,EPJA2000Bender},
and corresponds to keeping only the diagonal contributions in
Eq.~(\ref{e:cms}).
Solving the Schr\"odinger equation provides
the wave functions $\phi_q^i(\rv)$ and the
s.p.~energies $-e_q^i$ for the different
s.p.~levels $i$ and species $q$. 
We use in this work the standard nucleonic Skyrme force SLy4 \cite{sly},
but the results for hyperonic observables hardly depend on that choice.

\begin{figure}[t]
\vskip5mm
\includegraphics[width=59mm,angle=270,bb=100 220 360 220]{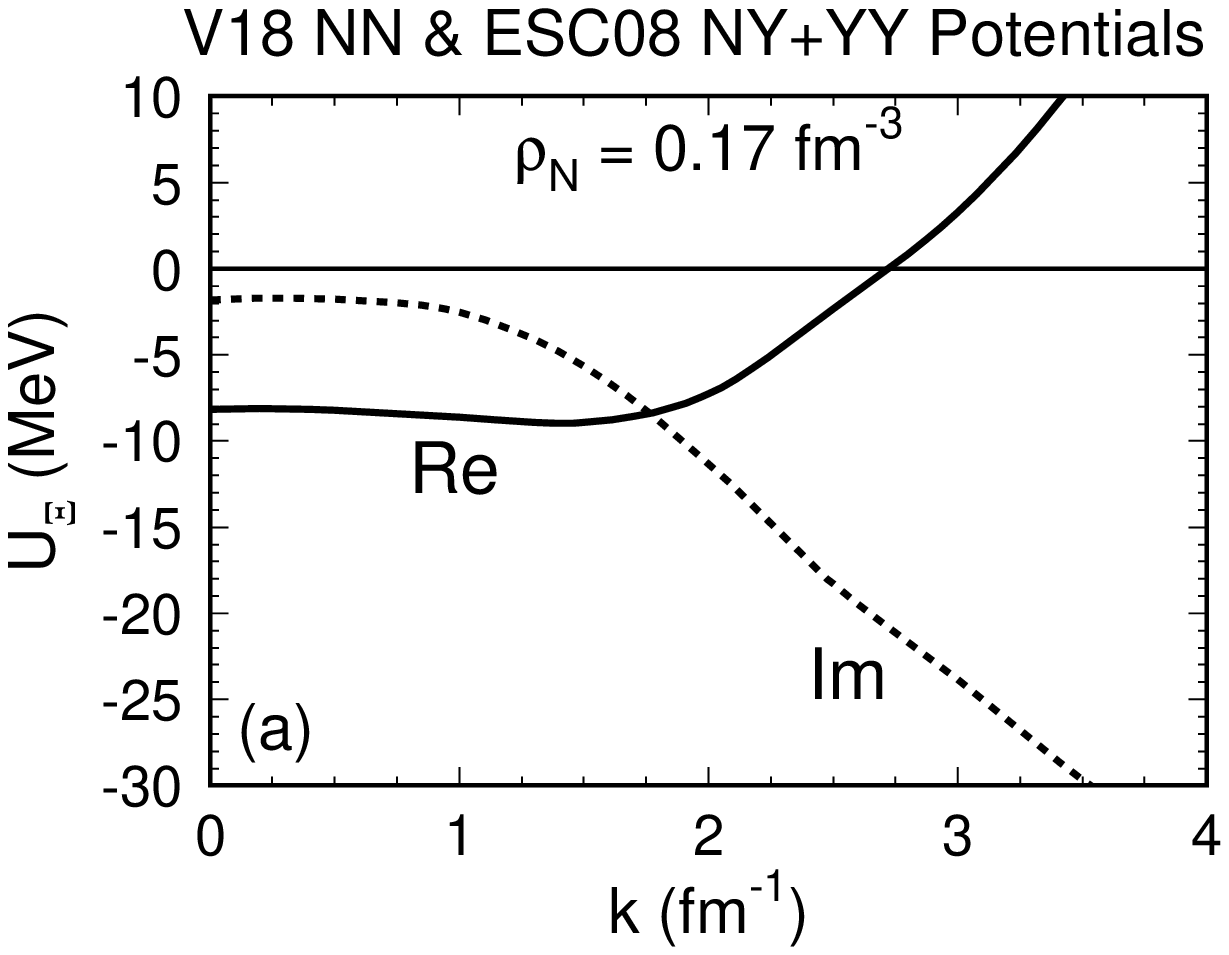}

\includegraphics[width=59mm,angle=270,bb=300 220 560 220]{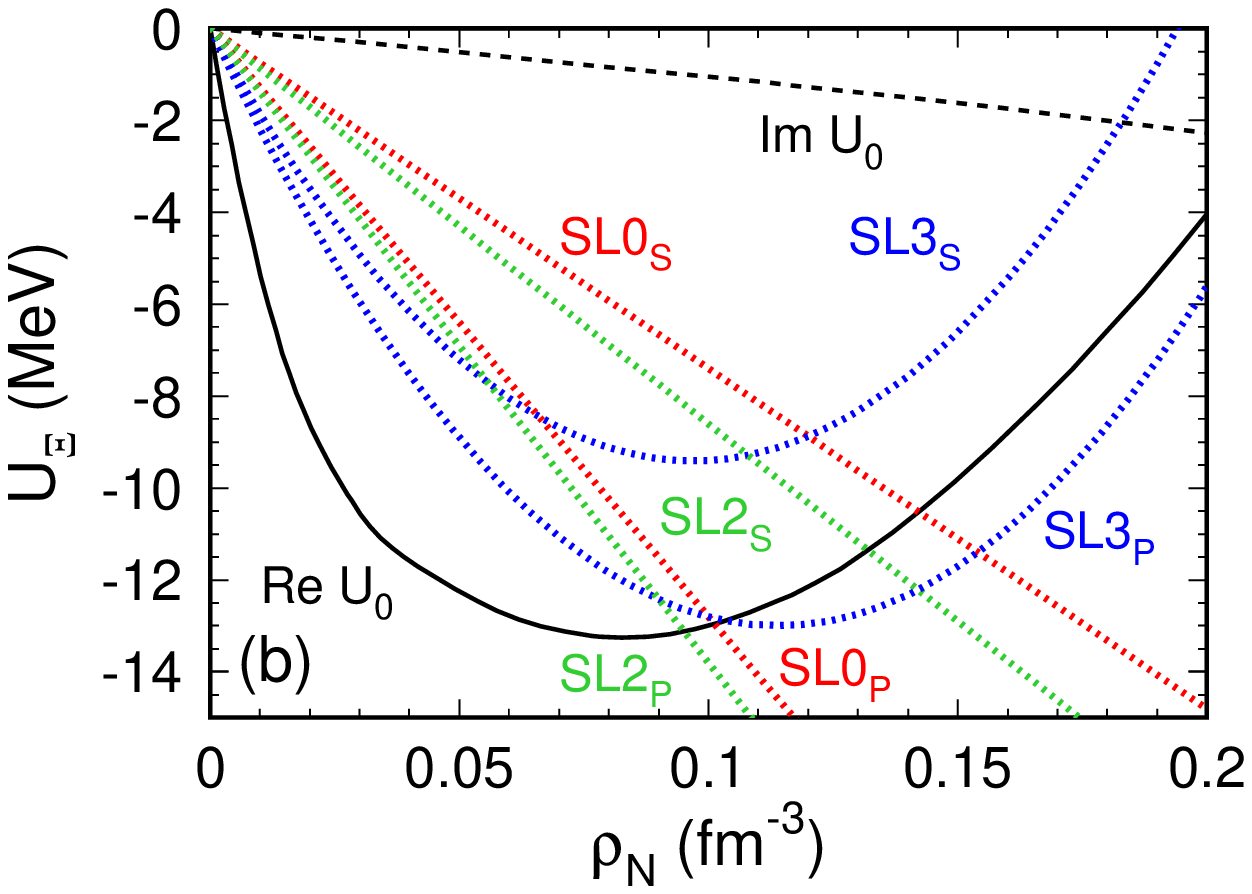}
\caption{(Color online)
Top:
BHF s.p.~potentials $U_\Xi(k)$
(real and imaginary parts)
in symmetric nuclear matter at $\rho_N=0.17\,\text{fm}^{-3}$,
obtained with the Nijmegen ESC08b $Y\!N$ model \cite{hypesc}.\\
Bottom:
Dependence on nuclear density of BHF s.p.~potential
$U_0\equiv U_\Xi(k=0)$
(black solid and dashed curves),
and the Skyrme SL0,2,3 mean fields $V_\Xi$
(dotted red, green, and blue curves)
in homogeneous nuclear matter,
Eq.~(\ref{e:vnm}).  The subscript $s (p)$ for each interaction denotes the potential obtained by the assumption of
$\Xi$ occupation in $s (p)$ orbit in Table I. See the captions to Table I for details.
}
\label{f:bhf}
\end{figure}

There are currently not enough data to determine all
$Y\!N$ interaction parameters $a_i$.
We therefore discuss three simple choices in the following:
The first, termed SL0,
is to consider only the volume parameter $a_0$.
This is justified by the fact that the $a_1$ parameter is related
to the hyperon effective mass,
Eq.~(\ref{e:mfit});
however, recent Brueckner-Hartree-Fock (BHF) calculations \cite{hypesc}
indicate that the \xim\ s.p.~spectrum is rather flat
and thus $m^*_Y/m_Y$ close to unity.
This is demonstrated in Fig.~\ref{f:bhf}(top),
where we plot the real and imaginary parts of the
momentum-dependent \xim\ BHF s.p.~potential $U_\Xi(k)$ \cite{hypesc}.
The imaginary part is fairly small, at least at low momenta,
$\Im U(0)/\Re U(0)\approx 0.2$,
which justifies to neglect it for the moment.
($\Im U$ depends strongly on the coupling of the \xin\ and $\lala$
channels, and the BHF results were obtained with the Nijmegen ESC08b
$Y\!N$ and $YY$ potentials.)

The parameter $a_2$ has no directly observable effect,
but determines the surface energy and might simulate finite-size effects
to some extend \cite{vaut}.
Motivated by the equivalent parameter of the
recently derived SLL4 $\Lambda N$ Skyrme force \cite{hypsky},
we introduce a further \xin\ Skyrme force, SL2, with the fixed value
$a_2=20\,\text{MeV\,fm}^5$,
and an adjustable $a_0$.

The parameter $a_3$ can be related to the nonlinear density
dependence of the \xim\ mean field in homogeneous nuclear matter,
\be
 V_Y(\ron) = a_0\ron + a_1\tau_N + a_3\rho_N^2 \:.
\label{e:vnm}
\ee
Again referring to the BHF results \cite{hypesc},
we fix this parameter roughly to
$a_3=1000\,\text{MeV\,fm}^6$,
and this force is termed SL3.
For comparison,
in the SLL4 $\Lambda N$ Skyrme force \cite{hypsky},
the equivalent optimal parameter is
$a_3^{\Lambda N}\approx700\,\text{MeV\,fm}^6$,
whereas $a_3^{NN}\approx {\cal O}(2000\,\text{MeV\,fm}^6)$
in typical nucleonic Skyrme forces \cite{shf}.

We show in Fig.~\ref{f:bhf}(bottom) the density dependence of the
SL0,2,3 Skyrme forces
(with parameters fixed in the next section)
in comparison with the BHF s.p.~potential
depth $U_\Xi(k=0)$ in nuclear matter.
One notes in particular the very different behavior of the SL0,2
and SL3 forces,
which has consequences for the predicted $\bxi$ values in light and heavy
hypernuclei later.
The volume term of the SL2 force has to provide more attraction than
that of the SL0 force,
because the SL2 surface term acts repulsive.

It is important to
stress that the Nijmegen (or any other) $Y\!N$ potentials do not provide an
independent {\em prediction} of the correct \xim\ mean field;
their parameter values have rather been adjusted motivated by different current
hypernuclear experimental data.
We use the BHF results here only in order to fix approximately
the value of the $a_3$ Skyrme parameter.

We discuss now the choice of the relevant parameter $a_0$
of the SL0,2,3 forces.
At the moment it is clearly premature to try to fix all \xin\ Skyrme parameters;
we use the different variations of the SL* force only
in order to investigate the qualitative physical consequences of the different
interaction terms in confrontation with the data.
For the same reason we do not introduce further parameters for
the isospin dependence of the interaction, e.g.,
\be
 a_0\rho_N \ra a_0^n \rho_n + a_0^p\rho_p \:,
\ee
in order to accommodate the Lane potential.
We will discuss later this possibility.

\begin{table}[t]
\caption{The \xim\ removal energies $\bxi$ (in MeV) of \xic\ and \xbe
 with the framework of (a) RMF and (b) SHF.
The calculated values in parenthesis are those in the case of switching off
$\Xi N$ Coulomb interaction.
The bold numbers have been fitted so as to reproduce the observed data of Kiso
event \cite{nakazawa}.
In  (a),  RMF$^{\sigma\omega\rho}$ ( RMF$^{\sigma\omega}$)
denotes results with  (without) the  isospin dependent potential,  adopting different  coupling constants
$\alpha_{\sigma\Xi}\equiv g_{\sigma\Xi}/g_\sigma$
and
$\alpha_{\rho\Xi}\equiv g_{\rho\Xi}/g_\rho$.
In  (b),
Results, SL0,SL2 and SL3 are obtained with  \xin\ Skyrme forces of
different  parameters $a_0$, $a_2$, and $a_3$,respectively. The subscripts $s$ and $p$
denote the orbit in which  $\Xi$ is trapped.}

\begin{ruledtabular}
\def\myc#1{\multicolumn{1}{c}{#1}}
\def\myr#1{\multicolumn{1}{r}{$#1$}}
\def\myrr#1{\multicolumn{2}{c}{$#1$}}
\renewcommand{\arraystretch}{1.2}
\begin{tabular}{lrrrrcr}
 (a) &\myrr{\alpha_{\sigma\Xi}} &\myr{\alpha_{\rho\Xi}}
  &\myc{$^{15}_{\hskip0.05em\Xi s}$C}
  &\myc{$^{15}_{\hskip0.05em\Xi p}$C}
  &\myc{$^{12}_{\hskip0.05em\Xi s}$Be} \\
\hline
 RMF$^{\sigma\omega\rho}_s$ & \myrr{0.295} & 1 & {\bf4.4}\ (1.1) &          & 1.7\ (-0.3) \\
 RMF$^{\sigma\omega}_s$     & \myrr{0.296} & 0 & {\bf4.4}\ (1.1) &          & 2.7\ ( 0.3) \\
 RMF$^{\sigma\omega\rho}_p$ & \myrr{0.313} & 1 & 9.4\ (5.7)      & {\bf1.1} & 6.1\ (~3.4) \\
 RMF$^{\sigma\omega}_p$     & \myrr{0.311} & 0 & 8.0\ (4.3)      & {\bf1.1} & 6.2\ (~3.4) \\
\hline
 (b) &\myr{a_0} &\myr{a_2} &\myr{a_3}
  &\myc{$^{15}_{\hskip0.05em\Xi s}$C}
  &\myc{$^{15}_{\hskip0.05em\Xi p}$C}
  &\myc{$^{12}_{\hskip0.05em\Xi s}$Be} \\
\hline
 SL0$_s$ & -74 &  0 &   0 & {\bf4.4}\ (0.9) &          & 2.4\ (-0.1) \\
 SL2$_s$ & -86 & 20 &   0 & {\bf4.4}\ (1.0) &          & 2.3\ (-0.2) \\
 SL3$_s$ &-194 &  0 &1000 & {\bf4.4}\ (1.1) &          & 2.6\ (~0.3) \\
 SL0$_p$ &-128 &  0 &   0 & 10.4\ (6.6)     & {\bf1.1} & 8.0\ (~5.2) \\
 SL2$_p$ &-138 & 20 &   0 & 10.0\ (6.2)     & {\bf1.1} & 7.3\ (~4.5) \\
 SL3$_p$ &-228 &  0 &1000 &  7.2\ (3.7)     & {\bf1.1} & 5.2\ (~2.6) \\
\hline
 \multicolumn{4}{l}{Exp.~or empirical data}
 & 4.38$\pm$0.25 & $1.11\pm0.25$ & $5\ (~2.2)$ \\
 \multicolumn{4}{l}{Ref.}
 & \myc{\cite{nakazawa}}
 & \myc{\cite{10LBe,private}}
 & \myc{\cite{khaustov,hiyama}} \\
\end{tabular}
\end{ruledtabular}
\label{t:bec}
\end{table}

\section{Results and discussion}
\label{s:res}

The calculated results,
$\Xi^-$ removal energies of \xic\ together with
the parameter values ($\alpha_{\sigma\Xi, \rho\Xi}$ and $a_{0,2,3}$, respectively)
are listed in Table~\ref{t:bec} within the frameworks of RMF and SHF.
We calculated the binding energies with and without Coulomb interaction
in order to see its effects.
For comparison, the $s$-state $\Xi^-$ removal energies for the
hypernucleus \xbe\ are also given.
In addition, the experimental data for \xic\ and \xbe\ are listed.
Especially, in the case of \xbe, we list empirical
data by the cluster model calculations \cite{hiyama},
assuming the observed $\Xi^-$ binding energy of \xbe\
in the ground state to be 2.2~MeV without Coulomb interaction.

\begin{figure}[t]
\vspace{2mm}
\includegraphics[scale=0.42]{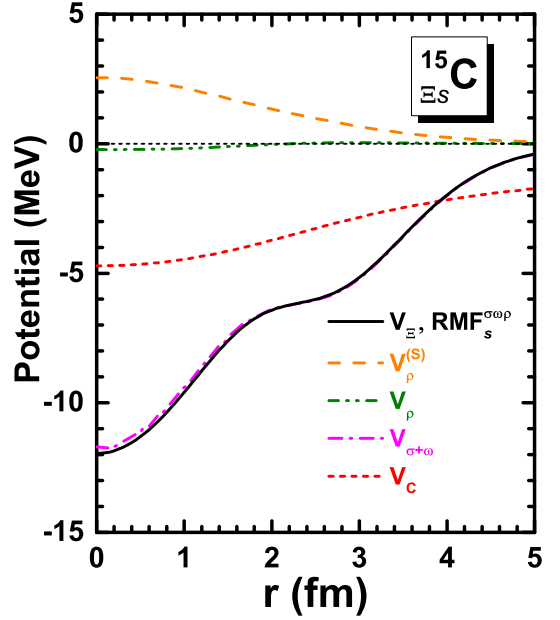}
\vspace{-3mm}
\caption{(Color online)
The potential between $^{14}$N and $\Xi^-$ obtained by  the RMF$^{\sigma\omega\rho}$ model.
The black solid line is shown with all components,  $V_{\sigma+\omega} + V_\rho +V_C$.
To see the contributions to $V_{\Xi}$, the self-coupling $V_\rho$ defined by $V_\rho^{(S)}$,
$V_\rho$,    $ V_{\sigma+\omega}$ and Coulomb potential $V_C$ are shown separately.}
\label{f:add}
\end{figure}

In the RMF calculations of \xic,
which has a pure isospin-zero nuclear core,
the entire $\rho$ field is generated by the hyperon
due to the hyperon self-interaction.
This $\Xi$ self-energy is considered as ``spurious''
and should be removed \cite{PRC1994Mares.49.2472}.
For the hypernucleus \xbe\ with non-zero isospin nuclear core,
this spurious field also exists.
In the following, as in Ref.~\cite{PRC1994Mares.49.2472},
we will isolate the $\Xi$-$\rho$ self-interaction by switching off
the $\rho$ coupling to the nucleons,
while the $\Xi$-$\rho$ interaction is left unchanged.
By comparing the results with those for $g_{\rho }=g_{\rho\Xi}=0$,
we obtain the spurious contribution of the hyperon self-interaction,
which we then subtract from the results of the full calculations.

For illustration, in Fig.~\ref{f:add}
the hyperon self-coupling potential $V_{\rho}^{(s)}$
and the different contributions to the local $\Xi^-$ mean-field potential
$V_{\Xi}$ in \xic\ are plotted with the force RMF$_{s}^{\sigma\omega\rho}$.
It can be seen that the spurious $\Xi$-$\rho$ potential is repulsive
with the central part around 2.5~MeV in \xic.
However, the potential $V_\rho$ contributed by the $\rho$ meson
is much reduced after subtracting $V_\rho^{(s)}$
and becomes very slightly attractive.
In the following, we will compare the results of the full
model RMF$^{\sigma\omega\rho}$
and a reduced model RMF$^{\sigma\omega}$ without $\rho$ meson,
in order to understand better the role of the
associated isospin dependence of the \xin\ interaction (Lane potential).
Thus, the isospin dependence has only a very weak effect of reduction by about 0.1~MeV,
comparing the results of the RMF$^{\sigma\omega\rho}$ and RMF$^{\sigma\omega}$
models.

As shown in Table~\ref{t:bec},
when we adjust the \xin\ interaction so as to reproduce
$\bxi=4.4$~MeV for \xic\ in the ground state
(entries with subscript ``$s$''), that is, \xis,
we find that $\bxi\approx1$~MeV without Coulomb interaction,
which means that the attraction effect of the Coulomb interaction is about 3~MeV.
With those \xin\ interactions,
the calculated $\Xi^-$ binding energy of \xbes\
is 1.7--2.7~MeV by the RMF and 2.3--2.6~MeV by the SHF model, respectively.
These energies are less bound in comparison with the ``empirical data''  of
$\bxi\approx5$~MeV with Coulomb.

Next, when we adjust the \xin\ interactions so as to reproduce
$\bxi=1.1$~MeV for \xic\ in the excited state
(entries with subscript ``$p$''), that is, \xip,
the calculated $\bxi$ of \xbes\ is 6.1--6.2~MeV by the RMF,
which is more consistent with the ``empirical value'' of $\bxi\approx5$~MeV.
Here, it should be noted that we have error bar, $\pm 0.25$ MeV, in $B_{\Xi^-}$ for $^{15}_{\Xi}$C.
Then, when we tune the  \xin\ interactions to be $0.86$ MeV and
$1.36$ MeV, which are upper and lowest   $B_{\Xi^-}$ for $^{15}_{\Xi}$C,
the energies of \xip are  5.41 MeV and 6.88 MeV, respectively, which are not away from the
'emprical data'.

Here, the $\Xi^-$ hyperon occupies the $1p_{3/2}$ state,
because the spin-orbit splitting of the $1p$ state is found very small,
0.06~MeV, and the $1p_{3/2}$ orbit is lower.
On the other hand, within the framework of SHF,
the calculated $\bxi$ using SL0, SL2, and SL3 are
8.0, 7.3, and 5.2~MeV, respectively.
In order to interpret the different results by RMF and SHF,
we will now discuss the associated potentials $V_{\Xi}$.

\begin{figure*}[t]
\centerline{
\includegraphics[scale=0.39]{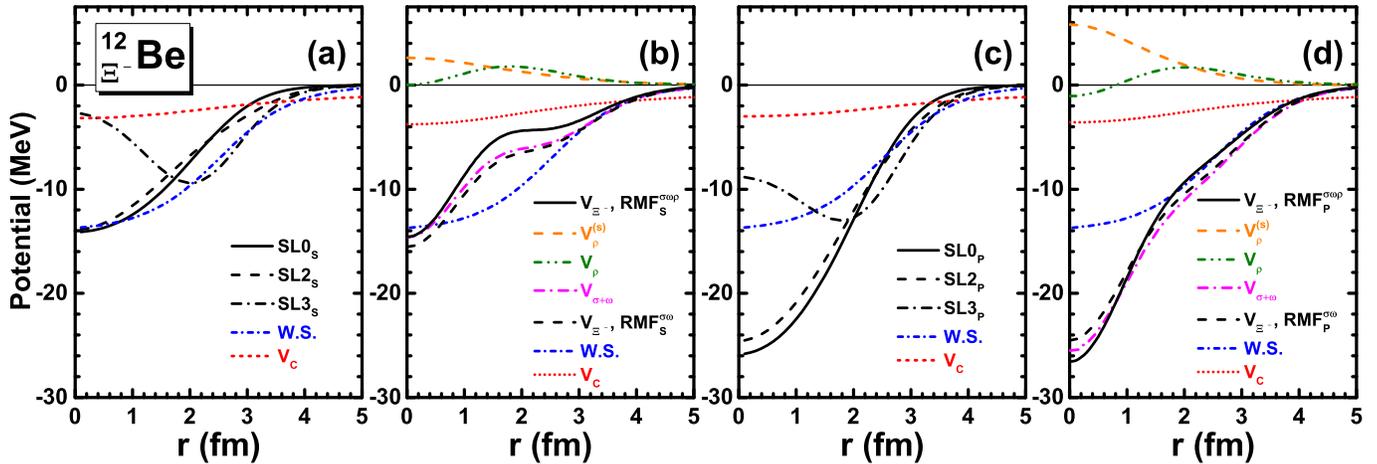}}
\vspace{-8mm}
\caption{(Color online)
Local \xim\  mean field $V_\Xi$ 
and Coulomb field $V_C$ (short-dashed red curves)
in \xbe,
obtained with the SHF (a, c) and RMF (b, d) models with  various
forces in Table I.
The Woods-Saxon mean field deduced in \cite{khaustov}
is shown for comparison by dash-dotted blue curve.
The panels (a) and (b) show $V_\Xi$ for the \xim\  occupation of $s-$orbit, while \xim\  occupies $p-$orbit in  (c) and (d).
 In (a) and (c),  results of different sets of Skyrme parameters SL0, SL2 and SL3 are shown by solid, dashed and dashed-dotted curves, respectively.
In the RMF cases (b) and (d), the components
$V_{\sigma+\omega}$ (long-dash-dotted magenta curve),
$V_\rho$ (dash-dot-dotted green curve),
and the spurious potential $V_{\rho}^{(s)}$ (dashed orange curve)
are  shown separately.  The sum of $V_{\sigma+\omega}$, $V_\rho$, $V_C$ and  $V_{\rho}^{(s)}$  is denoted by $V_\Xi$ RMF$^{\sigma\omega\rho}$ (solid curve),
while the sum without the isovector potential $V_\rho$ is given by $V_\Xi$ RMF$^{\sigma\omega}$ (long dashed curve).}
\label{f:v}
\end{figure*}

In Fig.~\ref{f:v},
the different RMF and SHF mean field potentials in the \xbe\ hypernucleus
are plotted,
including the local Coulomb field $V_C$,
the strong mean field $V_\Xi$ in Eqs.~(\ref{e:svt},\ref{e:vl})
together with the components
$V_{\sigma+\omega}=g_{\sigma\Xi}\sigma+g_{\omega\Xi}\omega_0$
and
$V_\rho=g_{\rho\Xi}\tau_{\Xi,3}\rho_{0,3}-V_{\rho}^{(S)}$
subtracting the spurious potential $V_{\rho}^{(S)}$
in the RMF case.
One notes in particular the very different shapes
of the SL0,2 and SL3 results,
caused by the different density dependence of those forces.
The shape of the RMF mean field corresponds roughly to the one of the
linear SL0,2 forces.
In the latter case the $a_2$ surface-energy term has the effect of widening
the potential well,
rendering the SL2 results more close to the RMF ones.

The Coulomb potentials obtained in the two frameworks are very similar.
For comparison also the WS mean field
($V$ and $r$ given in MeV and fm, respectively)
\be
 V_\text{WS}(r) = -14 / \left( 1 + \exp[(r-2.52)/0.65] \right) \:,
\ee
used in the analysis of Ref.~\cite{nakazawa} is shown in the figure.
In panels (a) and (b),
the depths of the mean fields SL0,2$_s$ and RMF$_s$
are nearly the same as the WS parametrization,
$V_\Xi(0)\approx-14\,\text{MeV}$.
However, the widths are much more narrow,
which provides less binding.
In panels (c) and (d),
the mean fields SL0$_p$ and RMF$_p$ are much deeper
than the WS parametrization,
remedying their narrow widths,
and finally provide much larger $\Xi^-$ removal energies.
It can be seen that the mean field SL0$_p$
has about the same depth,
but is wider than the one of RMF$_p$,
which leads to an about 2~MeV larger $\Xi^-$ binding energy
listed in Table~\ref{t:bec}.

Comparing with the ``empirical value'' of \xbes,
the interactions RMF and SL3 are capable to simultaneously reproduce
the data of \xicp\ and \xbes.
Therefore we support the claim that the ``Kiso event''  could
be an observation of the excited state in \xic, i.e., \xip,
which is consistent with one of the experimental interpretations.
With the compatible \xin\ interactions,
we then predict that the $\bxi$ of the ground state of
$^{15}_{\hskip0.05em\Xi s}$C
should be 8.0--9.4~MeV with RMF
and 7.2~MeV using SL3 with SHF.
This range is not small due to the fact that the $p$-state $\Xi$
probes mainly fairly low nuclear densities,
such that the behavior of the $\Xi$ mean field remains
largely unconstrained at normal nuclear density,
relevant for the $s$ state.

Furthermore, we have in this work only adjusted the isoscalar \xin\ interactions,
and disregarded fitting also the isospin dependence of the interaction.
This situation can only improve with the availability of more unambiguous and
precise data.

\section{Summary}
\label{s:end}

Motivated by the recent observation of the \xic\
$(^{14}\text{N}+\Xi^-)$ Kiso event,
which provides the first clear evidence
for a strongly bound $\Xi^-$ hypernuclear state,
we have studied the structure of that $\Xi^-$ hypernucleus
and the \xin\ interaction
within the framework of the RMF and SHF models.
The \xin\ interactions are constructed by reproducing the experimental data.
For the Kiso event,
we have two interpretations for the $\Xi$ binding energy $\bxi$,
$\approx4.4$~MeV or $\approx1.1$~MeV,
which could correspond to the ground state and excited state
of \xic\ with the $\Xi^-$ hyperon in the $1s$ and $1p$ orbits, respectively.

First, assuming \xic\ to be the ground state \xis,
and adjusting the \xin\ interaction so as to reproduce $\bxi=4.4$~MeV,
the calculated $\bxi$ of \xbe\ is 1.7--2.7~MeV with Coulomb interaction,
which is excluded due to the much smaller values of $\bxi$
than the empirical data $\approx5$~MeV,
unless the WS mean field is about 2--3~MeV less than 14~MeV for that nucleus,
or the Lane potential is unusually attractive for \xbe\ \cite{esc}.

Next, assuming \xic\ to be the excited state \xip,
we tune the \xin\ interaction to yield $\bxi=1.1$~MeV for \xicp,
the obtained \xbe\ becomes much more bound with respect to the
$^{11}$B$+\Xi^-$ threshold.
With Coulomb interaction,
it is 6.1--6.2~MeV by the RMF and 5.2~MeV by the SL3 of SHF,
which appear consistent with the empirical data $\bxi\approx5$~MeV.

Combining the above two cases,
the preferred interpretation of the Kiso event
is an observation of the excited state in \xic,
i.e., \xip\ by the approaches of RMF and SHF,
which is consistent with the experimental analysis.
Then the predicted $\Xi$ removal energy of \xic\ in the ground state is
7.2--9.4~MeV.
Currently, it is planned to perform an emulsion experiment to search for
double-strangeness hypernuclei at J-PARC this year.
Our prediction should be confronted with the future data,
which will also serve to constrain better the RMF and SHF \xin\ interaction
parameters.

\section{Acknowledgments}

We thank Y.~Yamamoto for stimulating discussions.
H.-J.~S.~thanks for the kind hospitality during his stay at RIKEN.
This work was partly supported by the JSPS Grant No.~23224006,
the RIKEN iTHES Project,
COST Action MP1304 ``NewCompStar'',
the Chinese Major State 973 Program No.~2013CB834402,
and the NSFC (Grants No.~11175002, 11335002, 11505157).


\end{document}